\documentclass[review]{elsarticle}
\usepackage{amsmath,amssymb,amsfonts}
\usepackage{algorithmic}
\usepackage{graphicx}
\usepackage[caption=false]{subfig}
\usepackage{textcomp}
\usepackage{xcolor}
\usepackage{color, soul}
\usepackage{import}
\usepackage[ruled,vlined]{algorithm2e}
\usepackage{setspace}
\usepackage[normalem]{ulem}
\usepackage{etoolbox}
\makeatletter
\makeatletter
\def\ps@pprintTitle{%
  \let\@oddhead\@empty
  \let\@evenhead\@empty
  \let\@oddfoot\@empty
  \let\@evenfoot\@oddfoot
}
\makeatother

\usepackage{hyperref}

\journal{International Journal of Critical Infrastructure Protection}

\begin{document}

\begin{frontmatter}

\title{Defensive Cost-Benefit Analysis of Smart Grid Digital Functionalities\tnoteref{mytitlenote}}

\author[hisaddress]{Jim Stright}
\author[hisaddress]{Peter Cheetham}
\author[hisnewaddress]{Charalambos Konstantinou\corref{mycorrespondingauthor}}
\cortext[mycorrespondingauthor]{Corresponding author}
\ead{charalambos.konstantinou@kaust.edu.sa}

\address[hisaddress]{Florida State University, Tallahassee, FL, USA}
\address[hisnewaddress]{King Abdullah University of Science and Technology (KAUST), Thuwal, Saudi Arabia}

\begin{abstract}
Modern smart grids offer several types of digital control and monitoring of electric power transmission and distribution that enable greater efficiency and integrative functionality than traditional power grids. These benefits, however, introduce greater complexity and greatly disrupt and expand the threat landscape. The number of vulnerabilities is increasing as grid-connected devices proliferate. The potential costs to society of these vulnerabilities are difficult to determine, as are their likelihoods of successful exploitation. In this article, we present a method for comparing the net economic benefits and costs of the various cyber-functionalities associated with smart grids from the perspective of cyberattack vulnerabilities and defending against them. The economic considerations of cyber defense spending suggest the existence of optimal levels of expenditures,  which  might  vary  among  digital  functionalities. We illustrate hypothetical case studies on how digital functionalities can be assessed and compared with respect to the costs of defending them from cyberattacks. 

\end{abstract}


\end{frontmatter}

{\section{Introduction}\label{sec:introduction}}

%
%
%
%
{Recognizing} the susceptibility of the U.S. power grid to cyberattacks, Congress  passed the Securing Energy Infrastructure Act, establishing a two-year pilot program to look at non-digital approaches for mitigating the potential effects of such attacks \cite{secact19}. Prior to the introduction in the late twentieth century of digital control and monitoring of the grid, cyberattacks on grid infrastructure had been only a theoretical concern. To date, disruptive power outages due to cyberattacks have been rare compared to outages from natural causes. Nevertheless, the likelihood that cyberattacks could severely curtail power availability is increasing as industrial control and monitoring of power grids becomes increasingly digitized \cite{mclaughlin2016cybersecurity}. 

Estimating the costs associated with rare but severe disruptions of power grids is difficult, whether these disruptions are due to cyberattacks or other causes. However, as existing grids evolve into smart grids with new types of functionalities as well as  vulnerabilities, it becomes increasingly important to try to understand, from a broad perspective, the costs as well as the benefits such technologies can be expected to introduce. 

Smart grid often refers to the shift in electrical generation from centralized to decentralized including the incorporation of energy storage, renewable generation, and bidirectional power flow. While this decentralization may be a defining goal for a smart grid \cite{konstantinou2020towards}, 
the process of integrating the enabling digital technologies occurs incrementally due to 
reliability and economic constraints of {the vast existing system of electricity generation and transmission}.
Many different digital technologies can or might be used to enable desired grid functionalities \cite{nordling2018social}. However, each of these technologies might introduce potential entry points for adversaries \cite{pate2018cyber, keliris2017ge}. Thus, it is of paramount importance to carefully consider how funding might best be allocated to defend desired grid functionalities against cyberattacks \cite{7436314}. Ideally, the allocation of limited cyber defense spending among various digitally-enabled grid functionalities should be applied before widespread integration of the enabling technologies within the system. 

Cybersecurity budgeting and spending practices often suggest prioritizing defense spending based on the criticality of digital assets but seldom explain how to determine the amount of security spending to apply to each asset. Existing methods for defending against digital information system cyberattacks (e.g., monitoring, discovering, patching vulnerabilities) usually assume potentially high but nevertheless bounded costs associated with successful attacks \cite{gordon2002economics, gordon2020integrating}. {R}eal-world incidents against cyber-enabled infrastructures, however, have the potential to cost not only potentially recoverable money but also irrecoverable lives. This makes a broad understanding of effective cyber defense vital. 
Building on the work of Gordon and Loeb \cite{gordon2002economics}, this work develops a framework for estimating the net economic benefits and costs associated with defending digitally enabled technologies against cyberattacks on cyber-physical energy systems \cite{9351954}.

\section{Assessing Digital Benefits and Costs}

In this article, the analysis of digital control and monitoring functions presumes as a base state the existence of a hypothetical functioning power grid which contains various digital technologies. These technologies are assessed in light of the most recent understanding of expenses and benefits associated with them, and in particular, the costs (and potential costs) associated with cyberattacks including defending against them. 

Central to our analysis is the concept of a grid digital functionality (GDF). A GDF is regarded as any digital technology, or any combination of digital technologies, that can be integrated into a power grid and potentially provide or enable a useful service to either an owning utility or grid customer (or both). The emphasis in identifying a GDF is based on its potential use in cyberattacks. In general, a given GDF might be an integral part of several types of cyberattacks. For example, residential smart meters which provide wireless digital residential power monitoring capabilities (considered as a  GDF) can reduce labor costs for utility companies, but they can potentially be used in several types of attacks, e.g., pricing cyberattack{s} and energy theft \cite{konstantinou2015cyber}. Strategically located synchrophasors within a grid can provide a useful monitoring capability for potential operator intervention \cite{silverstein2015value}. Such a monitoring capability is also considered a GDF, but distinct from an automated monitoring and control capability obtained by directly linking synchrophasors with certain flexible alternating current transmission system (FACTS) control devices via a digital communication channel. Such an automated combination of monitoring and control is considered a distinct GDF because it has potentially distinct vulnerabilities. Similarly, a digital control device which acts directly in response to analog inputs is regarded as offering yet another GDF. Another example includes residential wifi-enabled thermostats that provide a GDF of homeowner convenience but by so doing introduce yet another possible avenue for attacking the grid. 

{S}ome GDFs might be dependent on the presence of other GDFs. For instance, a GDF that permits control or monitoring of a substation condition, but only via digital commands from a control center,  
is dependent on the existence of that digital command functionality. {Indeed, many GDFs rely on the existence of fast, accurate, reliable and secure digital communication channels. The increasing incorporation of bidirectional power flows where once only unidirectional flows existed requires precise, near-real-time monitoring of power quality at many locations. The ability of utility customers to feed locally generated power back to the utility is enabled by the GDF of precise power quality measurements (via phasor measurement units), but this GDF depends on the existence of digital communication channels that are themselves GDFs. There are several types of digital communication channels, each with its own benefits and costs, and selection of appropriate communication channels is crucial to cost-benefit analyses of GDFs that depend on them. False data injection attacks and denial of service attacks that exploit weaknesses in digital communication channels can obviate any benefits of GDFs dependent on these channels. Highly secure, low latency channels are always desirable, but these qualities have associated costs} \cite{pal2020prekeying, 9430606, 9351954}. {In particular, successful cyberattacks against communication channels can eliminate the benefits of all GDFs that depend on them.} 

Let $F$ denote the set of all GDFs associated with a smart grid. For each $x \in F$ a method is  described for determining the expected net benefit that $x$ provides to the grid. The expected net benefit of $x$ is expressed as the increase in value provided by $x$ less the expected value of the costs and potential costs associated with $x$. 

It might be easiest to assign dollar values to digital technology benefits and expenses from the perspective of electric utilities. However, a more comprehensive approach would incorporate more general societal benefits and expenses in these valuations. For example, from their earliest deployments, residential smart meters have met stiff opposition from citizen groups on a number of grounds. 
Aside from concerns 
related to potentially hacked smart meters, many organizations have also noted privacy considerations and potential health risks due to radio frequency radiation, among other concerns \cite{wigan2012smart}. Actual worst-case costs associated with potential unintended consequences of smart meter deployment are difficult to estimate, as are the likelihoods that these consequences will be realized. 

For the GDF of smart meters, as well as many other GDFs, one of several possible ways of estimating such costs and likelihoods might be to convene a panel of experts with diverse backgrounds and task the panel to arrive at consensus values (under the direction, perhaps, of elected representatives). This could be done within a more general context of defending modern grids against many types of threats including, e.g., threats posed by natural or malicious electromagnetic pulses, direct physical attacks on grid components, disruption or elimination of GPS service, etc. A broad perspective like this might lead to contingency plans for cases where, despite best efforts, significant power outages develop and persist for extended periods of time.

A GDF can offer many potential benefits. The North American Synchrophasor Initiative (NASPI), for example, issued a report listing more than ten benefits offered by synchrophasors \cite{silverstein2015value}. It is safe to  assume that the electric utility industry is in the best position to calculate the benefits of GDFs, with the caveat that their calculations should be subject to appropriate public oversight. Given a GDF $x$, the total value of the benefits expected to accrue from its deployment is denoted as $\operatorname{\textit{Ben}}(x)$. Direct costs, denoted $\operatorname{\textit{DirCosts}}(x)$, associated with $x$, such as purchase and maintenance costs, is similarly  assumed to be calculated by the electric 
industry.

Other expenses associated with GDFs could often be more difficult to define and estimate than direct costs, and might be better estimated probabilistically. Here, we consider two types of such expenses based on whether those expenses are directly related to cyberattacks. 

Let $C$ denote the set of cyberattacks that can potentially adversely affect a power grid. Induced destruction of equipment, power disruptions, and theft of customer data will all be considered cyberattacks. As new digital technologies (such as Internet-of-Things (IoT) devices) interact with the grid, new attack possibilities are introduced, so the set $C$ should be expected to expand over time.

Let $N$ denote the set of non-cyber attack-related adverse events or cost impositions that might accrue due to inclusion of GDFs. These expenses can also be quite varied, including such things as storm-related damage to digital equipment, physical attacks on digital communication devices, and costs imposed by regulatory requirement compliance. 

Given any GDF $x \in F$, let $C_x \in C$ denote the attacks directly against $x$, i.e., that do not rely on  attacks on a different GDF. For each cyberattack $j \in C_x$, let $x_j$ represent an attack of type $j$ that uses the functionality $x$. Let $\operatorname{\textit{Cyb}}(x_j)$ denote the monetary value of a successful attack of type $j$ using $x$ together with the costs incurred attempting to defend $x$ against $j$. These defensive costs can be quite substantial. 
{Cyber defense spending is probably the most determinative factor in successful defense against cyberattacks, as the 2017 cyberattack in Ukraine seems to indicate.}
Typically, large companies spend from 3\% to 17\% of their information technology (IT) budgets on cyber defense, with 8\% to 10\% usually being adequate to ``achieve a lot of security'' \cite{clarke2019fifth}. A recent study found that financial services on average spend 10\% of their IT budgets on cybersecurity, approximately 0.2\% to 0.9\% of company revenue \cite{att}. One percent of the \$9.5 billion spent annually on IT by JPMorgan Chase is \$95 million \cite{chase}. Most companies, even very large ones, would probably prefer not to waste {even a single} percent of their IT spending, especially when it could be more effectively spent shoring up underfunded aspects of cyber defense. 
It is therefore important to try to determine a point at which useful cybersecurity spending becomes wasteful spending. 

Let $P(x_j)$ denote the probability that an attack $j \in C_x$ will succeed, given the grid's existing cyber defenses and skill levels of potential adversaries. Then the expected cost due to cyberattacks directly against $x$ is given by the summation $\sum_{j \in C_x}P(x_j)\operatorname{\textit{Cyb}}(x_j)$. 

Similarly, given a GDF $x \in F$ and adverse event $k \in N$, let $x_k$ denote the potential impact of $k$ on $x$. Let $\operatorname{\textit{Noncyb}}(x_k)$ denote the monetary value of $k$ on $x$. If the functionality $x$ is determined not to be relevant to $k$, e.g., if $x$ represents a software functionality and $k$ represents radio frequency interference, then $\operatorname{\textit{Noncyb}}(x_k)=0$. Let $P(x_k)$ denote the probability that the adverse event $k$ will impact functionality $x$. Then the expected cost due to non-cyber adverse effects is given by the summation $\sum_{k \in N} P(x_k) \operatorname{\textit{Noncyb}}(x_k)$. The expected net benefit ($\operatorname{\textit{ENB}}$) of the GDF $x$ can now be expressed as  

\begin{equation}
\begin{aligned}
\operatorname{\textit{ENB}}(x)&=\operatorname{\textit{Ben}}(x)-\operatorname{\textit{DirCosts}} s(x) \\
&
-\sum_{j \in C_{x}} P\left(x_{j}\right) \operatorname{\textit{Cyb}}\left(x_{j}\right) \\
&-\sum_{k \in N} P\left(x_{k}\right) \operatorname{\textit{Noncyb}}\left(x_{k}\right)
\end{aligned}
\label{eg:enb}
\end{equation}

The influence on $\operatorname{\textit{ENB}}(x)$ of expenditures on cyber defense can now be determined by noticing that only the terms $P(x_j)\operatorname{\textit{Cyb}}(x_j)$ appearing in the expression for $\operatorname{\textit{ENB}}(x)$ can be affected by such expenditures. Let $s_x$ denote expenditures intended to defend a functionality $x$ against cyberattacks. Some fixed value of $s_x$ is assumed to apply in the expression for $\operatorname{\textit{ENB}}(x)$. As $s_x$ increases, presumably the probabilities $P(x_j)$ of successful attacks against $x$ will decrease. However, the corresponding values $\operatorname{\textit{Cyb}}(x_j)$ will increase since these values incorporate the costs of cyber defense. Indeed, the summation of $P(x_j)\operatorname{\textit{Cyb}}(x_j)$  in the expression for $\operatorname{\textit{ENB}}(x)$ in Eq. \eqref{eg:enb} could be written 

\begin{equation}
f\left(s_{x}\right)=\sum_{j \in C_{x}} P_{s_{x}}\left(x_{j}\right) \operatorname{\textit{Cyb}}_{s_{x}}\left(x_{j}\right)
\label{Eq2}
\end{equation}

\noindent to more explicitly indicate its dependence on cyber defense spending. As a function of $s_x$, the $\operatorname{\textit{ENB}}$ of reintroducing functionality $x$ then becomes a constant minus $f(s_x)$

\begin{equation}
\begin{aligned}
&\operatorname{\textit{ENBCDS}}\left(s_{x}\right)= \operatorname{\textit{Ben}}(x)-\operatorname{\textit{DirCosts}}(x) \\
&-\sum_{k \in N} P\left(x_{k}\right) \operatorname{\textit{Noncyb}}\left(x_{k}\right)-f\left(s_{x}\right)
\end{aligned}
\label{eq:3}
\end{equation}

\noindent where $\operatorname{\textit{ENBCDS}}(s_x)$ is the $\operatorname{\textit{ENB}}$ of reintroducing functionality $x$ given the amount $s_x$ of cyber defense spending (CDS) intended to defend it. 

\begin{figure}[t]
\centerline{\includegraphics[width=0.7\textwidth]{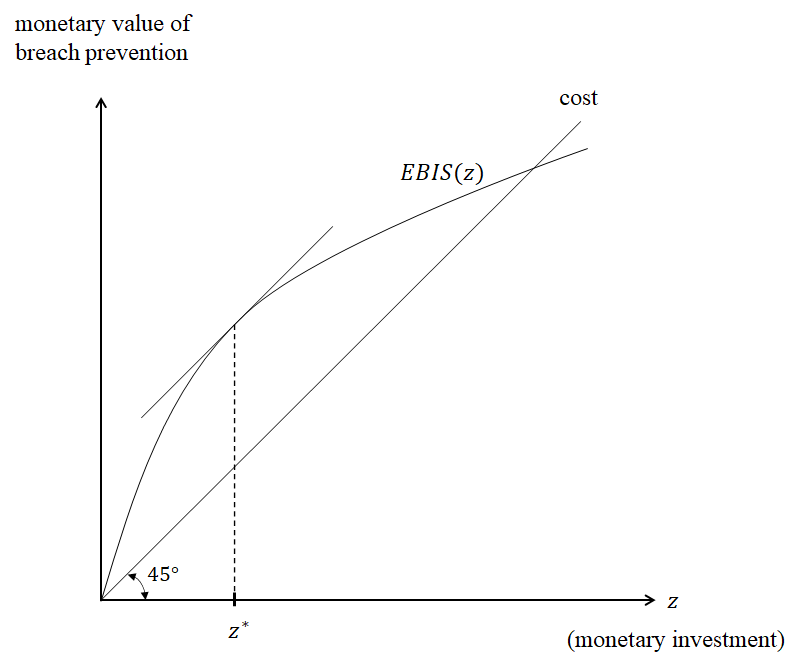}}
\caption{{Expected net benefit derived from expected benefit.}}
\label{fig:ebis_gordon}
\end{figure}

In \cite{gordon2002economics}, {Gordon and Loeb} derived, under a 
{carefully explained} set of assumptions, a similar function they designated $\operatorname{\textit{ENBIS}}$, meaning  $\operatorname{\textit{ENB}}$ from an investment in information security, which they demonstrated was strictly concave. {They obtained $\operatorname{\textit{ENBIS}}$ from the function $\operatorname{\textit{EBIS}}$ (expected benefit of an investment in information security) illustrated in} {Figure \ref{fig:ebis_gordon}.}

{The concave downward shape of EBIS was derived ultimately from the assumption that as investment in security increases, the probability of a successful breach decreases, but at a decreasing rate. Gordon and Loeb defined ENBIS, the expected net benefit of an investment in information security, as $ENBIS(z)=EBIS(z)-z$; that is, the expected benefit minus the cost of achieving it. The investment level $z^*$ in} {Figure \ref{fig:ebis_gordon}} {maximizes $ENBIS$; it is the level of investment which maximizes the expected benefit of the investment less its cost.} 

\begin{figure}[t]
\centerline{\includegraphics[width=0.66\textwidth]{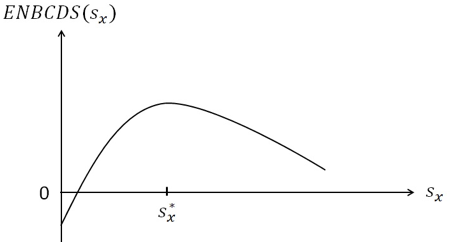}}
\caption{A generic expected net benefit {of} cyber defense spending ($\operatorname{\textit{ENBCDS}}$) function.}
\label{fig:genericENBCDS}
\end{figure}

A less rigorous but perhaps more intuitive appeal to the law of diminishing returns  
can illustrate why $\operatorname{\textit{ENBCDS}}${, as illustrated in} {Figure \ref{fig:genericENBCDS}}{, has the same downward-opening form as $EBIS$ (and $ENBIS$).}

Consider, for example, the situation where $x$ refers to the GDF of remote access to a power grid's Supervisory Control and Data Acquisition (SCADA) network. This functionality can enable more efficient control of the grid by legitimate power system operators but, as occurred in the 2015 attack on Ukraine's power grid \cite{zetter2016inside}, it also provides potential SCADA access to attackers. For this $x$, $s_x$ represents the expenditure on cyber defense efforts directly against illegitimate remote SCADA access. 

The mere introduction of remote SCADA access, with no cyber defense at all, clearly renders the grid more vulnerable than it had been. With zero spending on cyber defense, i.e., with $s_x=0$, the ``expected benefit'' of introducing remote SCADA access is actually an expected loss, i.e., $\operatorname{\textit{ENBCDS}}(s_x)<0$. Several cybersecurity measures can be and typically are instituted to try to thwart unauthorized access to a utility's SCADA network. A firewall can be installed to segregate the utility's corporate network from its SCADA network, but a firewall can be defeated by determined attackers. Various levels of multi-factor and user (and grid device \cite{zografopoulos2020derauth}) 
authentication can be employed to deny access to unauthorized personnel (e.g., demanding usernames and passwords with SMS two-factor authentication, then demanding physical ID cards, then requiring fingerprint scans), each successive level enhancing security, but never completely ensuring impenetrability. Additional measures could be taken, some entailing very significant additional expenditure (e.g., quantum authentication). Expenditures initially result in increased values of $\operatorname{\textit{ENBCDS}}(s_x)$, peaking at some value corresponding to an expenditure $s_x^*$, but beyond that level of spending, the spending itself (intended to thwart myriad remaining attack types in $J$) outweighs the expected gain in security. By analogy, after a homeowner has installed a certain number of different types of locks on a door, a would-be burglar will most likely look for a different door or a window.

\begin{figure}[t]
\centerline{\includegraphics[width=0.64\textwidth]{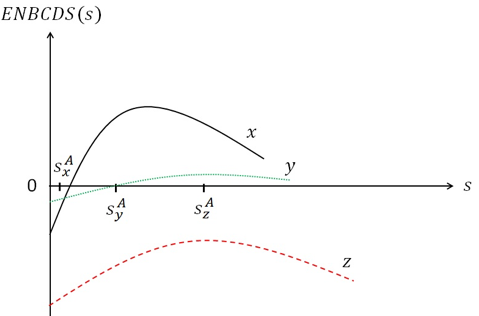}}
\caption{Comparing grid digital functionalities.}
\label{fig:compdig}
\end{figure}

Each GDF in $F$ has a concave $\operatorname{\textit{ENBCDS}}$ curve, and comparing curves for the GDFs in $F$ can provide insight into determining where best to allocate limited cyber defense spending. For example, Figure  \ref{fig:compdig} shows notionally three $\operatorname{\textit{ENBCDS}}$ curves for three hypothetical GDFs in $F$, labeled $x$, $y$, and $z$. In Figure  \ref{fig:compdig}, $s$ represents general cyber defense spending; $s_x^A$, $s_y^A$, and $s_z^A$ depict the actual levels of cyber defense spending directly on GDFs $x$, $y$, and $z$, respectively. Since $\operatorname{\textit{ENBCDS}}(s_z)<0$ always, it makes no economic sense to invest in $z$ at all, so $z$ should, with current understanding of that digital functionality, not be used. Economic considerations alone, however, might sometimes be overridden by other sociological factors. For instance, if $z$ represents the functionality of residential wifi-enabled thermostats, then $z$ is not a functionality that utilities themselves can eliminate. There may be insufficient political will to legislate against their use, or to hold responsible the manufacturers of wifi-enabled thermostats for their potential misuse, in which case the spending level $s_z^A$ (not necessarily borne entirely by utility companies) might be most appropriate in the sense that it minimizes the expected loss to society due to the use of $z$. As increasing numbers of consumer and industrial IoT devices like wifi-enabled thermostats and AC controllable loads impinge on power grids, 
increased attention to the likelihood of potential losses due to the types of cyberattacks they enable will become necessary\cite{pritchard}. 

If $z$ represents a digital functionality that could be replaced with an equivalent analog functionality that currently exists, or could economically be developed and deployed, then that analog functionality should be given consideration as a replacement for $z$. If $z$ represents a GDF that is inappropriate for the given market (e.g., power injections from residential solar systems to a utility that is abundantly supplied by geothermal sources) then $z$ should simply be recognized as an inapplicable functionality for that market. 

Notice that with the actual levels of spending on GDFs $x$ and $y$, no net expected benefit is realized. With any small increase in cyber defense spending for GDF $y$, some net expected benefit will be realized, which is not necessarily true for GDF $x$. However, if twice the actual spending level as currently exists for GDF $x$ is available to be added to the spending on any GDF, greater net expected benefit would be realized by applying that spending to GDF $x$ than to GDF $y$. 
Dependencies between the GDFs $x$ and $y$, if any, can also be important in interpreting Figure  \ref{fig:compdig}. For example, if the functionality $y$ is far more susceptible to successful attack if a prior attack against functionality $x$ has occurred, then it might be best to divert all spending on $y$ to $x$ and/or other GDFs, depending on their $\operatorname{\textit{ENBCDS}}$ curves and actual spending levels.

\section{Hypothetical Case Study: Unauthorized SCADA Remote Access}

In 2015, a successful cyberattack was accomplished  against the Ukrainian power grid by gaining access remotely to the grid's SCADA network \cite{zetter2016inside}. This mode of attack could not have been accomplished had remote access not been provided. Allowing remote access might create the following potential benefits for a similar hypothetical grid: 
allow increased response time to emerging grid conditions; reduce overall labor costs; 
and increase safety of utility personnel (e.g., during adverse weather events). Monetary values for these benefits and several others could be calculated by a utility company and added, giving the total benefit $\operatorname{\textit{Ben}}(x)$ the utility expects to realize from $x$, the GDF of remote SCADA access. 
Similarly, the utility can calculate the direct costs of instituting remote SCADA access. These might include costs for hardware and software, such as dedicated firewalls, monitoring systems and cell phones for employees, but only for those costs not directly in support of cybersecurity. These costs are denoted as $\operatorname{\textit{DirCosts}}(x)$.

Allowing remote access clearly opens a door for cyberattacks. In some installations, only monitoring of grid conditions can be accomplished, but in others, such as the Ukraine attack, surreptitious actors can perform certain control functions \cite{zografopoulos2021security}. Unauthorized access of any type, however, has the potential to be quite costly, and preventive measures are invariably taken to reduce the likelihood of such access. Some of those measures were mentioned above, but others exist as well, and the amount of spending on each will influence how effective the overall defense against unauthorized access will be. This level of spending was presented in Eq. \eqref{Eq2} where with increasing $s_x$ the probability $P_{s_x}(x_j)$ generally decreases and $\operatorname{\textit{Cyb}}_{s_x} (x_j)$ increases. 

A remote access capability probably introduces little additional cost due to non-cyberattack-related adverse events other than perhaps the cost due to potential failure of its implementing software due to the additional complexity it introduces. Perhaps these costs can be combined into a single term $P(x_k)\operatorname{\textit{Noncyb}}(x_k)$ where now the cost $k \in N$ is basically careful software engineering. The $\operatorname{\textit{ENBCDS}}$ could in this hypothetical case be expressed:

\begin{equation}
\begin{aligned}
&\operatorname{\textit{ENBCDS}}\left(s_{x}\right)=\operatorname{\textit{Ben}}(x)-\operatorname{\textit{DirCosts}}(x) \\
&-P\left(x_{k}\right) \operatorname{\textit{Noncyb}}\left(x_{k}\right)-f\left(s_{x}\right)
\end{aligned}
\end{equation}

\begin{figure}[t]
\centerline{\includegraphics[width=0.65\textwidth]{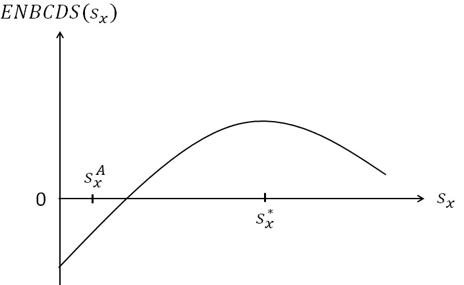}}
\caption{$\operatorname{\textit{ENBCDS}}$ for remote SCADA access.}
\label{fig:remoteSCADA}
\end{figure}

\noindent Depending on actual numerical values which could probably only be obtained through cooperation with affected utilities, $\operatorname{\textit{ENBCDS}}(s_x)$ might reasonably be represented by Figure  \ref{fig:remoteSCADA}.

It is conceivable that the level of spending in Ukraine used to protect against attacks using remote SCADA access was comparable to the level $s_x^A$ depicted in Figure  \ref{fig:remoteSCADA}. If additional cyber defense funding was available, some of it might gainfully have been used to approach the level $s_x^*$. If the optimal level of spending $s_x^*$ was prohibitively high, or if the shape of the curve $\operatorname{\textit{ENBCDS}}(s_x)$ was substantially different, the Ukrainian utilities might better have decided to forego remote SCADA access entirely, or at least control {functions enabled by remote} SCADA access. 

Among the cyberattacks in $C$ are those that rely on remote access to attack another GDF. In the case of the Ukrainian attack, for example, once the attackers gained SCADA network access, they attacked serial-to-ethernet converters used to process commands from the SCADA network to substation control systems \cite{zetter2016inside}. These converters provided the GDF of enabling remote re-closing of circuit breakers by grid operators following a blackout. The attackers remotely replaced firmware in the converters with malware to basically disable the converters. The utility dealt with the inability to remotely re-close breakers by dispatching personnel to re-close the breakers manually. Such manual fallback measures should be recognized as potentially necessary, and the costs associated with them should not be ignored when determining $\operatorname{\textit{ENBCDS}}(s_x)$. 

\section{Hypothetical Case Study: Smart Meter Deployment}

In this case study, we assume that all residential power metering in a hypothetical grid is accomplished using smart meters that can be wirelessly monitored and controlled. Such a metering infrastructure, when functioning as intended, has several benefits for residential customers as well as electric utilities.  
Some of the more important benefits include more detailed feedback to customers of usage (permitting possible customer savings); reduced number of outages; elimination of costly manual meter reading; and enablement of dynamic pricing by utility companies, reducing power production costs. In this hypothetical case, the benefit $\operatorname{\textit{Ben}}(x)$ of the GDF $x$ of smart meter deployment should probably explicitly include possible customer benefits, but again $\operatorname{\textit{Ben}}(x)$ should probably best be calculated by the utility company. 

A smart meter infrastructure entails several apparent direct costs and some significant ancillary or potential costs not directly related to cybersecurity. Direct costs might include the cost of the smart meters themselves; the cost of smart meter installation; the cost of IT to collect, store, and process metering data; and the costs of maintaining the smart meters and all their associated software. Such costs can be combined to yield $\operatorname{\textit{DirCosts}}(x)$. Among the potential costs not directly related to cybersecurity and which entail higher degrees of uncertainty are lawsuits arising from illnesses allegedly caused by smart meter radiation and from fires allegedly caused by malfunctioning smart meters \cite{meterlawsuits}; replacement of equipment due to storm-related damage; and costs of software upgrades, if needed, to facilitate demand response potentialities afforded by future deployment of internet-connected domestic appliances. Eventualities like these are included in the set $K$ and their combined expected cost is given by $\sum_{k \in N}P(x_k )\operatorname{\textit{Noncyb}}(x_k )$.

In the case of smart meters, in part because of their ubiquitous deployment, the set $C_x$ of cyberattacks directly against smart meters might be particularly significant. The types of attacks enabled by smart meter deployment include misuse of meter functionality by customers (e.g., for incorrectly reduced billing); unauthorized and perhaps malicious disconnection of selected customer service; theft, misuse, or corruption of collected customer data; and malicious widespread disconnection of customer service. 
These are among the attacks in $C_x$ that need to have probabilities of success associated with them. The combined expected cost of all cyberattacks is $\sum_{j \in C_x} P(x_j)\operatorname{\textit{Cyb}}(x_j)$.

\begin{figure}[t]
\centerline{\includegraphics[width=0.65\textwidth]{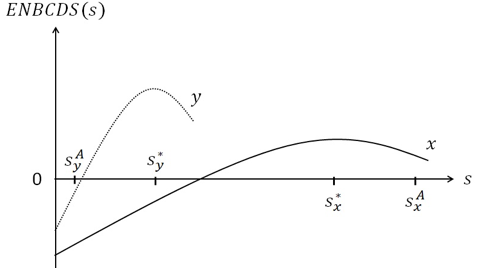}}
\caption{Comparison of hypothetical $\operatorname{\textit{ENBCDS}}$ curves.}
\label{fig:comparisonENBCDS}
\end{figure}

The $\operatorname{\textit{ENBCDS}}$ in this hypothetical case can be expressed as in Eq. \eqref{eq:3} where $f\left(s_{x}\right)$ is given by Eq. \eqref{Eq2} and with $P_{s_x}(x_j)$ generally decreasing and ${\operatorname{\textit{Cyb}}}_{s_x} (x_j)$ generally increasing as $s_x$ increases. 
The function $\operatorname{\textit{ENBCDS}}(s_x)$ might be represented by a curve similar to the curve labeled $x$ in Figure  \ref{fig:comparisonENBCDS}. This graph also contains a curve labeled $y$ which might represent the $\operatorname{\textit{ENBCDS}}$ on deployed digital protective relays. Some digital protective relays can be used without remote communication; the main cyber defense expense for these devices might be ensuring against resident malicious hardware or software implants. Intuitively, small cyber defense expenditures on digital protective relays might be expected to provide greater net benefit than the same level of expenditures on smart meters, which probably entail greater cyber vulnerability. If valid, the curves in Figure  \ref{fig:comparisonENBCDS} suggest that the amount $s_x^A-s_x^*$ of cyber defense spending should be diverted from smart meters to digital protective relays. Of course, many other possible GDFs are not depicted in Figure  \ref{fig:comparisonENBCDS}, and the levels of actual and optimal spending on each of them will need to be considered in any spending decisions. Nevertheless, Figure  \ref{fig:comparisonENBCDS} illustrates a potential benefit of trying to determine $\operatorname{\textit{ENBCDS}}$ curves for all grid digital functionalities actually employed or under consideration. 

\section{Conclusions}

Modern smart grids offer several types of digital control and monitoring of electric power transmission and distribution that enable greater efficiency and integrative functionality than traditional power grids. These benefits, however, introduce greater complexity and greatly disrupt and expand the threat landscape. The number of vulnerabilities is increasing as grid-connected  devices proliferate. The potential costs to society of these vulnerabilities are difficult to determine, as are their likelihoods of successful exploitation. In this article, we presented a method for comparing the net economic benefits and costs of the various cyber functionalities associated with smart grids from the perspective of cyberattack vulnerabilities and defending against them. The economic considerations of cyber defense spending suggest the existence of optimal levels of expenditures,  which  might  vary  among  digital  functionalities. Appropriate levels of defensive spending on each digital functionality should take into consideration the defensive requirements of all digital functionalities, the funding available for defensive measures, and the concerns of all electric power producers, suppliers, and consumers.  We illustrated hypothetical case studies on how digital functionalities can be assessed and compared with respect to the costs of defending them from cyberattacks.

\bibliography{mybibfile}

\end{document}